\documentclass[conference]{IEEEtran}
\IEEEoverridecommandlockouts
\usepackage{cite}
\usepackage{amsmath,amssymb,amsfonts}
\usepackage{algorithmic}
\usepackage{graphicx}
\usepackage{subfigure}
\usepackage{textcomp}
\usepackage{tabularx}
\usepackage[bottom]{footmisc}
\usepackage{xcolor}
\DeclareMathOperator*{\Max}{maximize}
\makeatletter
\newcommand{\linebreakand}{%
  \end{@IEEEauthorhalign}
  \hfill\mbox{}\par
  \mbox{}\hfill\begin{@IEEEauthorhalign}
}
\makeatother
\def\BibTeX{{\rm B\kern-.05em{\sc i\kern-.025em b}\kern-.08em
    T\kern-.1667em\lower.7ex\hbox{E}\kern-.125emX}}
\begin{document}

\title{Energy-aware Demand Selection and Allocation for Real-time IoT Data Trading\\
}

\author{\IEEEauthorblockN{Pooja Gupta}
\IEEEauthorblockA{\textit{School of Computer Science and Engineering} \\
\textit{The University of New South Wales}\\
Sydney, Australia \\
pooja.gupta@unsw.edu.au}
\and
\IEEEauthorblockN{Volkan Dedeoglu}
\IEEEauthorblockA{\textit{Data61, CSIRO}\\
Brisbane, Australia \\
volkan.dedeoglu@csiro.au}
\and
\IEEEauthorblockN{Kamran Najeebullah}
\IEEEauthorblockA{\textit{Data61, CSIRO}\\
Brisbane, Australia \\
kamran.najeebullah@csiro.au}
\linebreakand
\IEEEauthorblockN{Salil S. Kanhere}
\IEEEauthorblockA{\textit{School of Computer Science and Engineering} \\
\textit{The University of New South Wales}\\
Sydney, Australia \\
salil.kanhere@unsw.edu.au}
\and
\IEEEauthorblockN{Raja Jurdak}
\IEEEauthorblockA{\textit{School of Computer Science} \\
\textit{Queensland University of Technology}\\
Brisbane, Australia \\
r.jurdak@qut.edu.au}
}

\maketitle

\begin{abstract}
Personal IoT data is a new economic asset that individuals can trade to generate revenue on the emerging data marketplaces. Typically, marketplaces are centralized systems that raise concerns of privacy, single point of failure, little transparency and involve trusted intermediaries to be fair. Furthermore, the battery-operated IoT devices limit the amount of IoT data to be traded in real-time that affects buyer/seller satisfaction and hence, impacting the sustainability and usability of such a marketplace. This work proposes to utilize blockchain technology to realize a trusted and transparent decentralized marketplace for contract compliance for trading IoT data streams generated by battery-operated IoT devices in real-time. The contribution of this paper is two-fold: (1) we propose an autonomous blockchain-based marketplace equipped with essential functionalities such as agreement framework, pricing model and rating mechanism to create an effective marketplace framework without involving a mediator, (2) we propose a mechanism for selection and allocation of buyers' demands on seller's devices under quality and battery constraints. We present a proof-of-concept implementation in Ethereum to demonstrate the feasibility of the framework. We investigated the impact of buyer's demand on the battery drainage of the IoT devices under different scenarios through extensive simulations. Our results show that this approach is viable and benefits the seller and buyer for creating a sustainable marketplace model for trading IoT data in real-time from battery-powered IoT devices.

\end{abstract}


\section{Introduction}
Personal IoT devices generate an overwhelming amount of sensing data, which could be utilized for solving real-time problems within smart cities, transportation, and waste management. However, this data is usually accumulated in private silos that are difficult to access and remain under-utilized due to the lack of incentives and data sharing mechanisms. Given the potential value of this data, the notion of a data marketplace has been proposed, which facilitates the trade between data owners and potential buyers.

Conventional centralized approaches based on the client-server model \cite{b1,b2}, are built around a trusted third party that conducts all management operations and controls every aspect of a trade. Hence, raising issues such as single point of failure, lack of scalability, the need for expensive infrastructure and limited privacy. Recently, decentralized data marketplaces based on blockchain technology \cite{b3} have gained popularity due to the technology's salient features including transparency, security, anonymity and immutability. Blockchain has the potential to overcome the aforementioned problems by distributing the computation across multiple nodes. It provides a verifiable and auditable ledger by recording all transactions, ensures integrity and trust in every transaction using the consensus mechanism and facilitates owner-controlled access and data sharing.

Existing data marketplace approaches that use blockchain provide specific functionalities. For instance, in \cite{b4}, the blockchain platform provides access control policies that enable data owners to manage who can access their data and for how long. In \cite{b5}, blockchain is used as a distributed data catalogue for listing details of offerings in a reliable and immutable manner. Although these approaches allow the sharing of IoT data, they do not have all the essential functions of a marketplace. A functional and effective marketplace must support all the buying/selling activities such as listing and discovering data, managing contracts, pricing, billing and reputation mechanisms for entities to rate each other. 

Furthermore, participants' economic utility is an essential consideration for usability and wider adoption of the marketplace. Existing work focusing on economic aspects adopted incentive mechanisms based on pricing \cite{b6,b7}. These pricing strategies can motivate users to contribute their data. However, to guarantee the long-term engagement of participants, an economic model is required that analyzes the trade-off between maximizing the seller's and the buyer's utility.

The utility of a seller depends on the revenue generated from data trading while that of the buyer depends on getting quality data for the desired rate and duration. However, IoT devices are restricted in their capabilities and derive their energy from batteries or external sources using harvesters. Recharging or using harvesting techniques can prolong the battery lifetime. However, typically, the available energy is insufficient to stream data continuously to multiple buyers in real-time, impacting seller's and buyer's utility. For instance, consider an application developer providing location-based services by purchasing real-time location data from multiple sellers. The developer specifies his desired data quality in terms of accuracy and latency, sampling interval, and duration. These parameters directly impact the battery consumption of the seller's device. Existing techniques \cite{b8} for improving power consumption of IoT devices usually compromise data quality for energy efficiency. High-quality data(high accuracy and frequency, low latency) usually consumes more energy. For instance, the positioning approach which provides the most accurate location information (GPS: 6m, Assisted GPS: 60m, Cell-ID/Wifi Positioning System: 1600m), consumes the most power (GPS: high, AGPS: medium, Cell-ID/WPS: low) \cite{b9}. 

Similarly, when an IoT sensor works at a low sampling interval and for a longer duration, it imposes a heavy workload on different components (processor, network, storage and sensors) of the device, draining the battery rapidly. As an example, if a smartphone owner sells his location data by sending frequent (low sampling interval) data, the phone battery depletes more quickly. This degrades the smartphone user experience and impacts other buyers' utility with whom the seller has already made an agreement. On the other hand, if a low data rate setting is applied, the data quality is reduced, degrading the buyer's utility. Due to the costs associated with continuously sensing and transmitting data, a seller has to make an optimal decision to serve buyers' demands in real-time based on the devices' current residual energy and capabilities. Thus, it is vital to find a solution that maximizes the participants' utilities while considering the limited capabilities and constraints for creating a sustainable marketplace.

This work follows on from our earlier proposal for an IoT data marketplace, where we presented a 3-tier architectural design. The approach suggested in \cite{b10} was based on a distributed P2P broker for providing discovery and matching functionalities. Additionally, data subscription and register smart contracts were collectively used to automate data trading. This paper presents an enhancement of our prior marketplace design that addresses the two challenges mentioned above: (i) satisfying the seller's and buyer's utilities when trading real-time sensing data from battery-operated IoT devices, and (ii) offering key mechanisms for a marketplace, i.e., agreement management, pricing model, and reputation mechanism. We realize these functionalities leveraging smart contracts to record agreement details, participant's ratings and data prices in blockchain in a transparent, secure, efficient and automated way without involving any mediator. Our paper makes the following contributions:

\noindent $\bullet$ We present details of the blockchain-enabled framework including the architectural design, the participants, and their interactions.

\noindent $\bullet$ We formulate the demand selection and allocation optimization problem to maximize the seller's revenue by taking into account the resource-constrained nature of IoT devices and solve it using our novel greedy heuristic algorithm. Simulation results confirm that both sellers and buyers will benefit from the proposed algorithm.

\noindent $\bullet$ We implement key components of the marketplace such as agreement framework, pricing model and rating model using smart contracts and present a proof-of-concept implementation of our framework in Ethereum testnet to demonstrate its feasibility.

The rest of the paper is organized as follows. In Section II, we survey the related research. Section III presents an overview of the marketplace. In Section IV, we present integer linear formulation of the optimization problem followed by the details of our smart contracts in Section V. Proof of concept implementation and evaluations are presented in Section VI, and Section VII concludes the paper.

\section{Related work}
In this section, we discuss existing works which can be broadly categorized into two groups: task allocation optimization and blockchain-based data marketplace frameworks.

\noindent\textbf{Task allocation optimization} - Demand selection and allocation are essential for reducing IoT device energy consumption and increasing seller's revenue generation. \cite{b11} presents a mixed-integer linear programming formulation of assigning services to multiple heterogeneous network interfaces in an IoT device. The problem aims to minimize the total cost of utilizing the interfaces' resources while satisfying all the services' requirements in contrast to our case of selecting a subset of demands. In \cite{b12}, the authors maximize the profit with consideration of the task's urgency based on execution time deadline and QoS constraints to efficiently allocate resources and reduce the penalty cost due to agreement violations. However, we aim to select and allocate demands which require completion of the demand without any disruption irrespective of the execution time deadline. \cite{b13} presents a systematic literature review of various resource allocation methods in the IoT context. It concludes that collaboration of resource modeling, allocation, and monitoring is necessary to enable proper and continuous operation of IoT resources. We considered these guidelines while designing our framework.

\begin{table*}[t!]
  \begin{center}
    \caption{Comparison of blockchain based marketplace solutions}
    \label{table2}
    \begin{tabularx}{18cm}{|X|X|X|X|X|X|X|X|}
    \hline
      Elements & \cite{b4} & \cite{b14} & \cite{b5} & \cite{b15} & \cite{b16} & \cite{b17} & \cite{b18}\\
        \hline
        Smart contract application & Access control & Access control & Data offering, receipt & Registration, data offering & Data offering & Access control & Access control, data integrity\\
        \hline
        Data transfer & Broker & P2P & MQTT & Swarm & SDPP\cite{b19} & MQTT & pub-sub\\
        \hline
        Real-time support & Yes & No & Yes & No & Yes & Yes & Yes\\
        \hline
        Select \& match & No & DHT & browse blockchain & browse blockchain & browse blockchain & No & No\\
        \hline
        Pricing model & No & No & No & No & No & No & No\\
        \hline
        Data Agreement  & No & No & Yes & No & No & No & No\\
		\hline
		Rating model & No & No & No & Rate quality of the data & Rate quality of the data & No & Rate quality of the data\\
      \hline
    \end{tabularx}
\end{center}
\end{table*}

\noindent\textbf{Blockchain-based marketplace} - Table~\ref{table2} compares the recent blockchain-based marketplace frameworks to identify the supported functionalities with respect to real-time IoT data trading. The main criteria for our comparison are the smart contract application in realizing marketplace functions, data transfer method, real-time support, discovery and selection mechanism, pricing model, data agreement, and rating model. Through this comparison, we demonstrate the need for a holistic and effective marketplace design equipped with these key marketplace functionalities. Our framework supports (i) selection and matching based on available battery in devices to maximize seller's and buyer's utilities, (ii) the formation of customized and flexible agreement based on users' specifications, (iii) dynamic data pricing to encourage a seller to provide high-quality data, (iv) tuning of reputation scores based on the malicious behavior of dishonest actors and (v) data transfer over secured a TCP connection. 

\section{Overview of marketplace}
In this section, we present the blockchain-enabled framework for trading IoT data in real-time in an autonomous and decentralized manner, the main actors and their roles, and the high-level interactions to facilitate data trading.

\subsection{Real-time IoT data marketplace}

\begin{figure}[b!]
\centerline{\includegraphics[width=\linewidth]{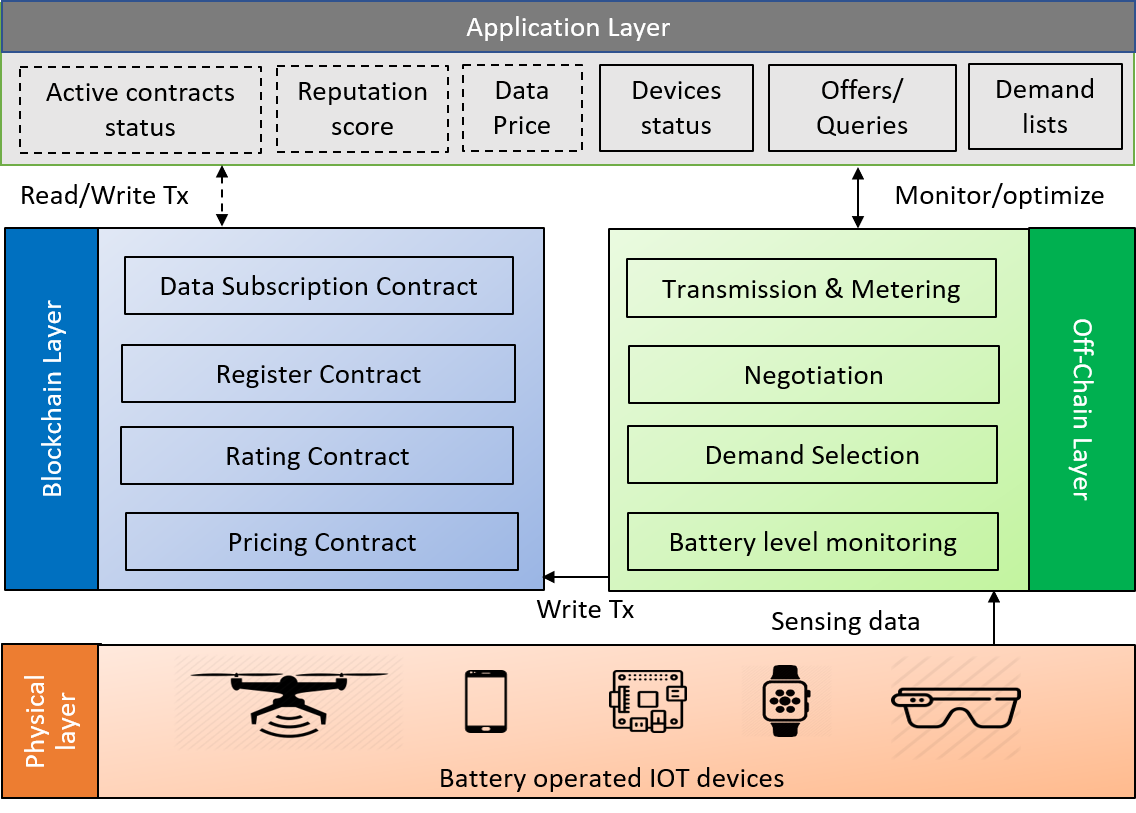}}
\caption{Multi-layered Blockchain Framework.}
\label{fig2}
\end{figure}

Our IoT data trading framework has four layers: physical, blockchain, off-chain and application stacked as shown in Figure~\ref{fig2}. Due to space limitations, some of the technical implementation details are omitted. We assume that the seller has at least one computationally capable device to host this framework, which will serve as the gateway for other resource-constrained devices that together form the physical layer.

The physical layer consists of all the battery-operated IoT devices such as sensors, drones, smartphones, watches, etc. As per the buyers' requirements, the sensors acquire data and transfer them to the gateway periodically.

The off-chain layer performs activities that are not recorded on the blockchain. We assume that following are some of the critical components needed for resource-constrained devices: 

\noindent (a) The battery monitoring entity periodically polls the devices to get the current battery status which plays an important role in device allocation to demands.

\noindent (b) The demand selection component selects and allocates received demands to the seller's devices, while satisfying the buyers' requirements and maximizing the seller's revenue using an optimization module explained in Section IV.

\noindent (c) The negotiation component uses Contract Net Protocol \cite{b20} for negotiating the terms between seller and buyer until an agreement is reached

\noindent (d) The transmission and metering component performs the data transfer via a secure TCP connection established between the buyer and the gateway device. The metering sub-component maintains the count of the transferred data samples used in the settlement.

At the blockchain layer, we utilize four smart contracts (namely: data subscription, register, pricing, and rating contracts) to facilitate the essential functionalities of the marketplace. The write transactions invoke these smart contracts at different stages of the data trading process either from the off-chain or the application components.  Register and data subscription contracts collectively provide an agreement framework to ensure the integrity of an agreement, non-repudiation of the owners and guarantee that the participants' behaviors automatically conform to the terms of the agreements. The pricing contract supports a competition-based price model to assess the price of the IoT data based on market dynamics. The rating contract records the trading history and uses it to evaluate the reputation score of the actors and adds safeguards by tuning reputation scores based on malicious behaviors such as collusion, false feedback, contract violations, etc. The details of the smart contracts, their methods and associated transactions are given in Section V.

\begin{figure*}[ht!]
\centerline{\includegraphics[width=0.7\linewidth, height=6cm]{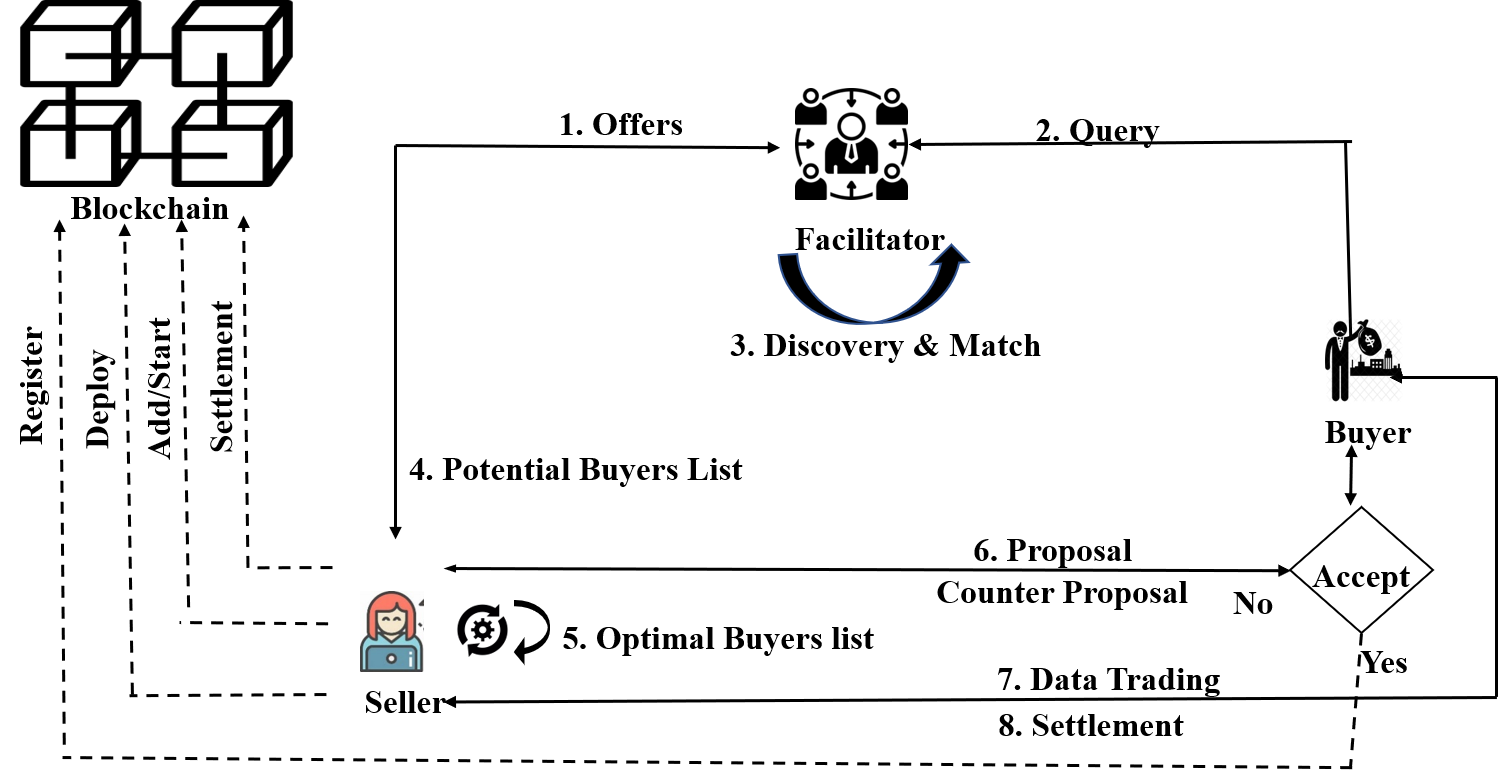}}
\caption{High-level interactions of actors in marketplace}
\label{fig1}
\end{figure*}

\subsection{Actors and their roles}

Our model involves three rational actors with specific roles linked to their participation in the data marketplace.

\noindent $\bullet$ \textit{Seller} - A seller uses two strategies to generate a sustainable income: (1) maximizing revenue by using the available resources on IoT devices as efficiently as possible, and (2) attracting buyers by offering add-on features such as high-quality IoT data or sensitive data which are distorting, revealing or intruding data owner's privacy. The seller posts the offerings to the facilitator as metadata. The metadata consists of information such as a public identifier (i.e., changeable public key) to anonymize the identity of the seller, list of IoT devices owned, geographical co-ordinates associated with the sensing data, temporal context of the data which can be static or dynamic and data usage license which defines the appropriate use of the data and also includes terms for restrictions of reselling of the data. In this work, we do not address the problem associated with reselling the data by unauthorized agents. Digital watermarking techniques \cite{b21} could be used to solve this problem.

\noindent$\bullet$ \textit{Buyer} - The buyer's requirements are classified into context-based requirements such as data type, location, temporal context, and seller's rating and quantitative-based requirements such as data quality, sampling interval and duration. The difference between the two is that the latter is dependent on the battery availability of the seller's devices while the former is independent of available battery.

\noindent$\bullet$ Facilitator - To address the scalability issue, we leverage geographically distributed facilitators that are interconnected in a P2P network and form an overlay network spanning the entire globe. Each facilitator is expected to oversee a particular service area/geographic region. We assume that the facilitator is a trusted and highly resourced entity whose motivation is to receive incentives in return for its service. All the sellers and buyers must register with a facilitator nearest to their geographical location and create a profile. 

\subsection{High-level interactions}

The end-to-end interactions among the actors are illustrated in Figure~\ref{fig1}. First, a facilitator receives queries and offerings from registered buyers and sellers respectively. Next, the facilitator performs semantic matching between offerings and queries using ontological models \cite{b22} and creates a list of potentially matching buyers and sends it to all identified sellers. The seller on receiving the potential list of buyers' demands uses optimization, as explained in section IV, to identify desirable buyers whose queries can be satisfied based on the battery availability of his devices. Next, he sends the data offerings directly to the desired buyers, followed by an optional negotiation process. Finally, the different stages of trading such as agreement formation, settlement, etc. are performed using smart contracts. Data is transferred off-chain over a secured TCP connection to buyers.
\section{Demand selection and allocation mechanism} 
The selection of demands discussed in Section III involves two steps. First, the facilitator performs the matching and selection based on the buyer's contextual requirements. Note that, quantitative-based requirements depend on the battery availability, a fast-changing factor, of the seller's device. To avoid network overheads associated with the seller updating the facilitator with the current battery level of his devices, a facilitator performs matching based only on the context-based requirements. Second, the selection is made based on the quantitative requirements at the seller end. The demand selection component of the off-chain layer performs this task intending to maximize the seller's revenue and meet the quality requirements and ensure completion of the selected demands without any interruptions. This mechanism of energy-aware demand selection and allocation is referred to as EDSA.

\begin{figure}[t!]
\centerline{\includegraphics[width=0.8\linewidth]{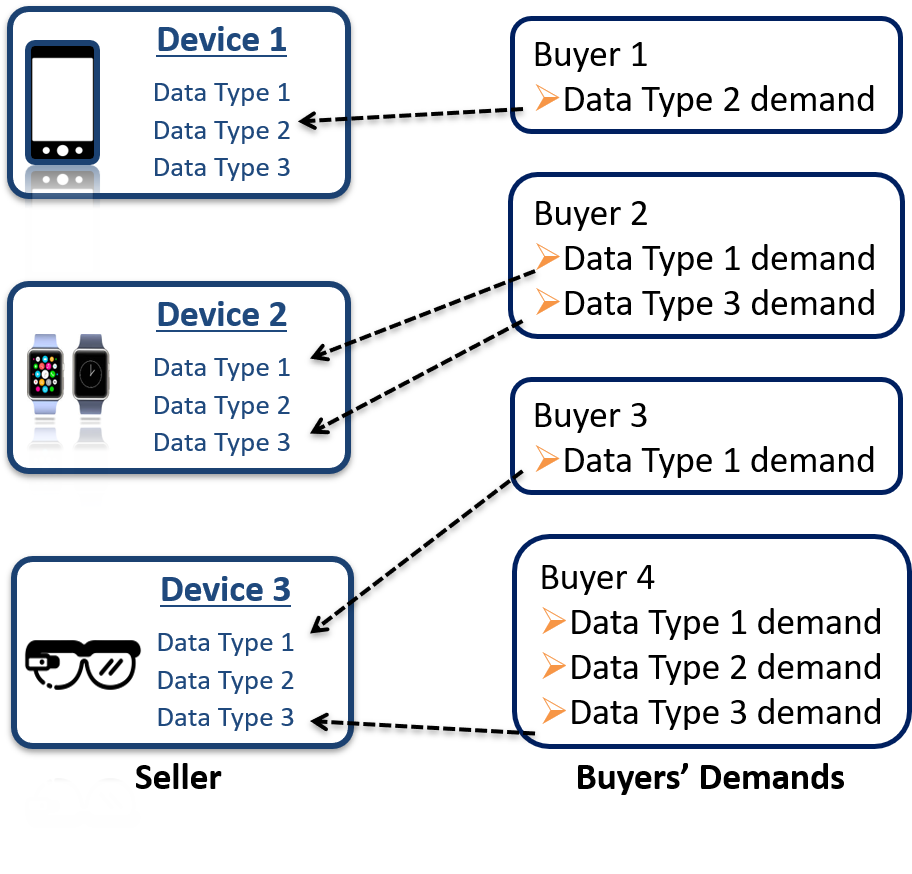}}
\caption{EDSA sample scenario}
\label{fig_sampleScenario}
\vspace{-0.5cm}
\end{figure}

In this section, we formally state the EDSA problem. Given a set of demands where each demand is expressed in terms of data type, sampling interval, duration, and quality, the aim is to find a subset of demands that maximize the seller's revenue while satisfying the battery, quality and allocation constraints. Figure~\ref{fig_sampleScenario} describes the EDSA sample scenario for a seller with three devices offering three different data types and four buyers with different data demands. All the demands of $buyer_1$, $buyer_2$, and $buyer_3$ are selected and assigned to $device_1$, $device_2$, and $device_3$, respectively, while only one demand of $buyer_4$ is selected and assigned to $device_3$.

The EDSA problem is formulated as an Integer Linear Program (ILP) as outlined below. Suppose a seller owns $D$ resource-constrained devices with battery levels $B_i$, for $i=1,2,\dots,D$. Each device provides a set of data types $S$. Without loss of generality, let $Q_{ij}$ be the highest quality threshold that a \textit{device i} can meet for a \textit{data type j}. There are a total of $C$ potential buyers, where each buyer may have multiple demands with different data types requirements. The unit data price $P_{ij}(q^k)$ is defined on the basis of the \textit{device i}, \textit{data type j} and the quality demand $q^k$ of the $buyer_k$. $N_{j}^k=(d_{j}^k/s_{j}^k)$ is the total number of samples of \textit{data type j} demanded by $buyer_k$, where $d_{j}^k$ is the duration and $s_{j}^k$ is the sampling interval. We assume that energy consumption for sensing varies for different sensors, while energy consumption for processing and transmitting data is same for different data types. The energy consumption $E_{ij}^k$ of \textit{device i} is the energy required for sensing, processing, and transmitting \textit{data type j} to the $buyer_k$ as given in \cite{b23}. 

The following ILP formulation outputs a selection of demands that optimally maximizes the total revenue while meeting the battery, quality and allocation constraints. The objective function is the total revenue generated by the demands selected and allocated to an IoT device of the seller. The EDSA problem can be formulated as:

\begin{align}
    &\Max_{x_{ij}^k \in \{0,1\}}
    \begin{aligned}[t]
       &\sum_{i}^{D}\sum_{j}^{S}\sum_{k}^{C} x_{ij}^k N_{j}^k P_{ij}(q^k)
    \end{aligned} \notag \\
    &\text{subject to} \notag \\
    &\quad \sum_{k}^{C}\sum_{j}^{S} x_{ij}^k E_{ij}^k \leq B_{i},\   \   \forall i, \label{batteryConstraint}\\
    &\quad \sum_{i}^{D} x_{ij}^k \leq 1,\   \   \forall k,\forall j, \label{allocationConstraint}\\
    &\quad x_{ij}^k q^k \leq  Q_{ij}  ,\   \   \forall i, \forall j,\forall k \label{qualityConstraint}
\end{align}

where the decision variables $x_{ij}^k =1$ if $data$ $type_j$ demand of $buyer_k$ is assigned to $device_i$, and $x_{ij}^k =0$ otherwise. Eq. \ref{batteryConstraint} captures the battery constraint of the devices. The allocation constraint in Eq. \ref{allocationConstraint} ensures that each buyer demand is served by at most one device of the seller. Eq. \ref{qualityConstraint} represents the quality constraint of the devices which defines the assignment restrictions.

Decision making in the EDSA can be regarded as the Multiple Knapsack Problem with Assignment Restrictions (MKAR) \cite{b24}. MKAR is considered in \cite{b24} in which the profit and weight of each item are the same. A more generic problem, in which profits and weights are different, was studied by \cite{b25}. It uses linear relaxation to find optimal vertex solutions followed by solving an instance of reduced MKAR problem iteratively. MKAR is NP-hard in the strong sense and no simple method to run on resource-constrained devices such as dynamic programming is known to apply to the MKAR. Also, given the low latency demand of a marketplace, computing an optimal solution may not be ideal. Therefore, we use a modified greedy heuristic approach \cite{b26} to include quality constraints to solve the EDSA.

\hrulefill

\textbf{Algorithm}: NR-based algorithm executed by the seller

\hrulefill

\begin{description}
  \item [1:]Initialization of the Demand list.
  \item [2:] \textbf{for each} dem in Demand
  \item [3:] \quad Calculate NR
  \item [4:] \textbf{end for}
  \item [5:] Sort demands in descending order of NR
  \item [6:] \quad \textbf{if} elements in NR are equal:
  \item [7:] \quad \quad Sort demands in descending order of prices 
  \item [8:] \quad \textbf{end if}
  \item [9:] end sort
  \item [10:] Sort devices in ascending order of battery availability
\item [11:] \textbf{for each} dem in Demand
\item [12:] \quad \textbf{for each} dv in device
\item [13:] \quad \quad Compare Device battery $\geq $ demanded energy
\item [14:] \quad \quad \quad Compare Device quality $\geq $ demand quality
\item [15:] \quad \quad \quad \quad Assign dem to dv
\item [16:] \quad \quad \quad \quad Update the available battery of dv
\item [16:] \quad \quad \quad \quad \textbf{break}
\item [18:] \quad \textbf{end for}
\item [19:] \textbf{end for}
\end{description}

The intuition behind our algorithm is that selecting demands with high prices and low energy consumption may result in high total revenue for the seller under energy constraints. Based on this intuition, we define the \textit{Normalized Revenue} (NR) of demand as the ratio of the generated revenue to the energy consumption of the demand. If the prices are chosen according to the costs, this scheme also maximizes the seller's total profit. Our NR-based algorithm starts by initializing the list of demands (Line 1) and calculating the NR values for each demand in the list (Line 2-4). Then, the list of demands is sorted in descending order of NR values (Line 5-9). Note that, demands with equal NR values are further sorted based on their prices. The rest of the algorithm (Line 11-19) assigns demands to devices using the sorted demands list (starting from the demand with the highest NR value), checking the battery availability and device quality constraints and updates the available battery of the devices. 

\section{Marketplace Elements}

After identifying the desired list of buyers' demands using the NR-based algorithm, the seller sends the data offer directly to the buyers. If both parties reach an agreement, a smart contract encoded with terms of the agreement is compiled into bytecode and deployed on the blockchain with a unique address. Transactions or messages are used to invoke the smart contract to update or modify its state. These transactions can be either unisig or multisig. Unisig transactions require the signature of either buyer or seller, whereas multisig transactions require the signature of both buyer and seller. For example, a transaction corresponding to the registration of the subscription contract between buyer and seller is a multisig transaction requiring the signature of both entities.

A smart contract can consist of several functions. So, the application binary interface (ABI) is required to specify which function in the contract to invoke along with the address of the contract. Subscription details, participant's ratings, and data prices are recorded in the blockchain using data subscription, rating, and pricing contracts, respectively. The ABIs and the transaction details for different contracts are given below.

\noindent\textbf{Data Subscription Contract (DSC)} : The terms of agreements are encoded in unisig DSC. Each seller-buyer pair have a single DSC to provide flexibility and customization based on their specifications. The DSC maintains the list of subscriptions which includes subscription identifier, seller identifier, buyer identifier, device identifier, data type, start time, periodicity, duration, quality score ($QS$), privacy risk score ($RS$), total cost, payment granularity, transaction status, negotiation information. \textit{Device id} is the hash of the device MAC address to ensure the seller's claim of owning the device and the information is stored in hash form for the sake of privacy. The DSC provides the following ABIs \cite{b10} to manage the subscription list: 

\noindent $\bullet$ \textit{subscriptionAdd}(): This method adds a new subscription request to the existing subscription list. The transaction $Tx_{add}$ can only be issued by the seller. $ID_b$ is the buyer's id and $Hash_{data}$ is the hash of the subscription details. $Sig_{s}$ and $PU_{s}$ are the signature and public key of the seller respectively.

\begin{equation}\label{tx_add}
Tx_{add} = [ID_{b} | Hash_{data} | Sig_{s} | PU_{s}]
\end{equation}

\noindent $\bullet$ \textit{subscriptionInfo}(): This ABI receives the subscription id and returns the corresponding subscription details. The transaction $Tx_{info}$ can be initiated by either of the concerned parties where $SID$ is the subscription id, $Sig$ and $PU$ are the signature and public key of either seller or buyer.

\begin{equation}\label{tx_info}
Tx_{info} = [SID | Sig | PU]
\end{equation}

\noindent $\bullet$ \textit{subscriptionStart}(): Subscription is started by the seller at the subscription start time using $Tx_{start}$.

\begin{equation}\label{tx_start}
Tx_{start} = [SID|Sig_{s}|PU_{s}]
\end{equation}

\noindent $\bullet$ \textit{subscriptionSettlement}(): This ABI is executed by $Tx_{settle}$ where $D_{count}$ is a positive integer while $F$ is a real number in (0,1). Both seller and buyer are required to submit the data count $D_{count}$ and provide feedback $F$ to the other actor based on his experience. This ABI performs settlement by comparing the received data counts and taking the following action based on both parties' data counts. Case 1: If data counts from both buyer and seller are equal, i.e., no conflicts exist, an invoice is generated for the buyer and payment is released to the seller. Case 2: Otherwise, in a dispute situation, the actor with a higher reputation score, managed by a rating contract explained later, is trusted.

\begin{equation}\label{tx_settle}
Tx_{settle} = [SID | D_{count} | F | Sig | PU]
\end{equation}

\noindent $\bullet$ 
\textit{subscriptionDelete}(): On completion of the subscription, $Tx_{del}$ is used to delete the subscription entry from the subscription table.

\begin{equation}\label{tx_delete}
Tx_{del} = [SID|Sig_{s}|PU_{s}]
\end{equation}

\noindent \textbf{Register Contract:} This contract maintains a DSC look-up table comprising of the following fields: contract id, creator id, seller public key, buyer public key, DSC address, ABIs name. The register contract uses the functions \cite{b10} discussed below to manage the look-up table:

\noindent $\bullet$ \textit{contractCreate}(): A newly deployed DSC can be added to the contract lookup table using multisig $Tx_{create}$  where $DSC_{address}$, $DSC_{ABI}$, and $ID_{s}$ are the DSC address, DSC ABI , and the the seller ID, respectively.

\begin{equation}\label{tx_create}
Tx_{create}  = [DSC_{address}|DSC_{ABI}|ID_{s}|Sig_{b}|PU_{b}|Sig_{s}|PU_{s}]
\end{equation}

\noindent $\bullet$ \textit{contractRemove}(): This ABI deletes the contract entry from the lookup table. It subsequently performs the \textit{SelfDestruct} operation resulting in removal of the storage and code from the state \cite{b30}. It is invoked by multisig $Tx_{remove}$ where $CID$ is the contract ID.

\begin{equation}\label{tx_remove}
Tx_{remove}  = [CID|Sig_{b}|PU_{b}|Sig_{s}|PU_{s}]
\end{equation}

\noindent $\bullet$ \textit{contractGet}(): $Tx_{get}$ is issued by either of the parties to get the DSC contract address and the associated ABIs.

\begin{equation}\label{tx_get}
Tx_{get} = [CID|Sig|PU]
\end{equation}

\noindent \textbf{Pricing Contract}: A pricing contract maintains a ledger with the following fields: timestamp, data type identifier, price, quality score and privacy risk score. A message $Mx_{price}$ is issued to record the price of the traded data type whenever a new DSC is deployed or a new subscription is added. $ID_{type}$ is the data type identifier, $Hash_{data}$ is the hash of the price details and $Sig_{contract}$ is the signature of the issuing contract.

\begin{equation}\label{Mx_price}
Mx_{price} = [ID_{type}|Hash_{data}|Sig_{contract}]
\end{equation}

We have adopted a competition-based pricing model in our framework. This pricing strategy is suitable for highly competitive markets, such as ours, which have a large population of individual sellers interested in selling their IoT data in return for incentives. In a competition-based pricing model, the seller can follow the market going rate charged by other individuals selling the same data type. This strategy does not require any complex computation and dynamically varies based on the market trend. The competitor's price information $Price_{d,t}^i$ recorded in the ledger for $i^{th}$ transaction is used to calculate the price index $P_{d,t}^i$ of \textit{data type d} at a certain \textit{time t} using Eq. \ref{index_price}. $QS_{d,t}^i$ and $RS_{d,t}^i$ are quality score and privacy risk score respectively. Then, base price $Price_{base}^d$ is calculated by averaging $P_{d,t}^i$ for the number of \textit{transactions S} in a given \textit{time interval T} using Eq. \ref{base_price}.

\begin{equation}\label{index_price}
P_{d,t}^i= \frac {Price_{d,t}^i}{QS_{d,t}^i + RS_{d,t}^i + 1}
\end{equation}

\begin{equation}\label{base_price}
Price_{base}^d = \frac {\sum_{i=1}^{S} P_{d,t}^i}{S}, t \in T
\end{equation}

However, in a highly competitive scenario, price is not a major differentiating factor to draw the buyers. Therefore, the seller must provide value-added features as discussed in section IIIB. We quantify these value-added features by the Quality score ($QS$) and privacy risk score ($RS$). Quality Score is the weighted average of the buyer's demand for quality levels and preferences. The privacy risk score of data is calculated using the risk matrix technique proposed in \cite{b27}. The price of the data is proportional to the scores mentioned above. Moreover, smart contract related execution fees ($P_{exe}$) can also be divided between the seller and buyer as per the negotiated terms. Our pricing model evaluates the price of a particular \textit{data type d} using Eq. \ref{price} where $\beta$ is the agreed-upon share of $P_{exe}$. 

\begin{equation}\label{price}
    Price=(1+QS +RS)Price_{base}^d + \beta P_{exe}
\end{equation}

\begin{figure*}[t!]
\centerline{\includegraphics[width=0.7\linewidth, height=6cm]{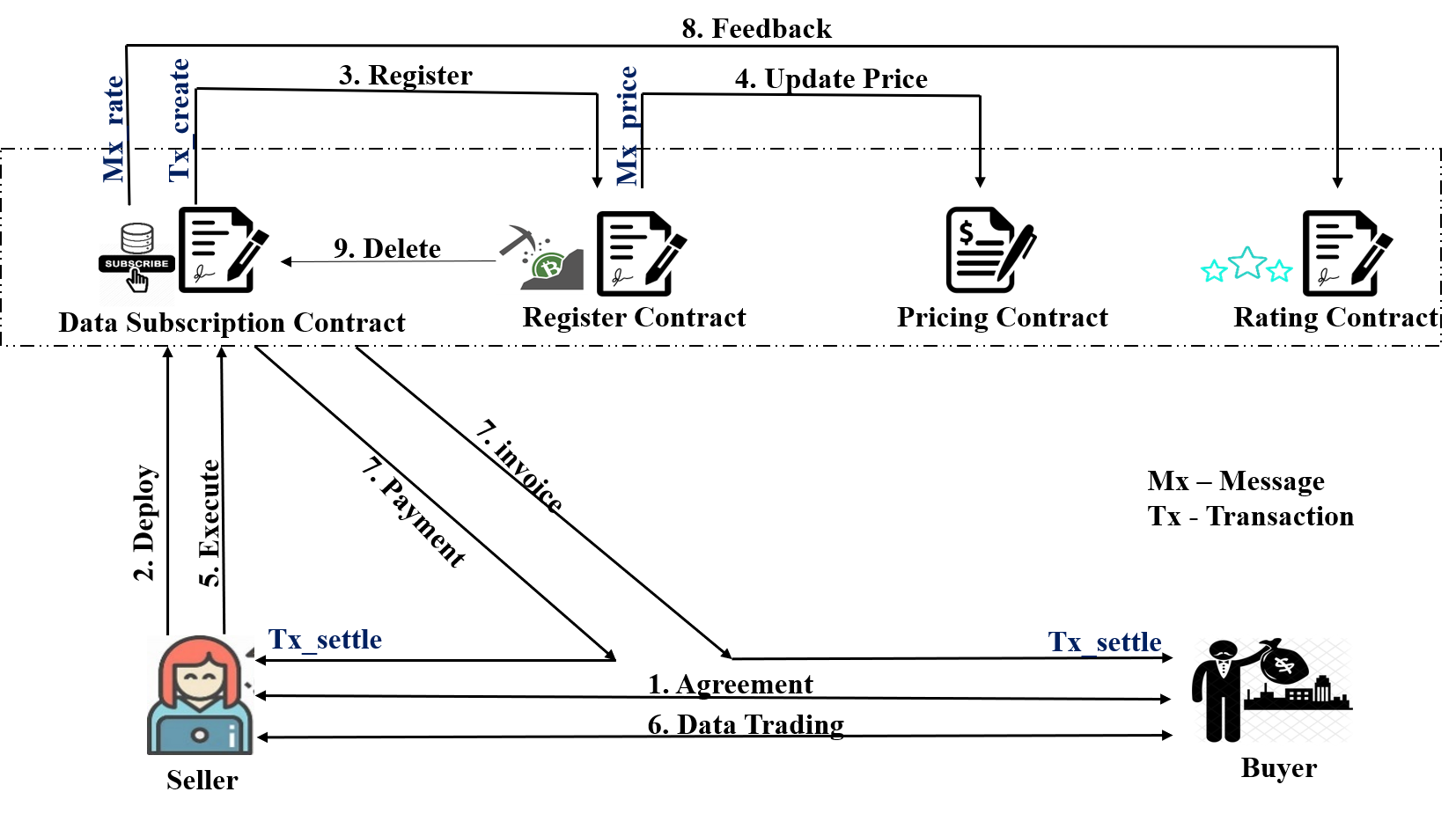}}
\caption{Autonomous data trading using smart contracts}
\label{smart_contract}
\end{figure*}

\noindent \textbf{Rating contract} - In the marketplace, actors should be trustworthy as economic interests are relevant. Reputation plays a crucial role in C2C (consumer to consumer) trading to build trust, facilitate a smooth transaction and reduce risk \cite{b28}. The rating contract calculates the reputation score based on the actor's trading history in the marketplace. Post-settlement, DSC issues a message $Mx_{rate}$ to update the trade history between \textit{actor i} and \textit{actor j}. Feedback $F_j$ and $F_i$ are received by \textit{actor j} and \textit{actor i} respectively. $T_{value}$ is the transaction value of the current trade. The rating contract records following fields in the ledger: maximum transaction value in the trade history ($v_m$), total positive feedback received ($F_j^+$), total feedback received, total number of failed contracts $c_{fail}$ due to agreement violation, time of last request ($ToLR$) and the total number of transactions $T_{i,j}$ between \textit{actor i} and\textit{ actor j} in a very short span of time from $ToLR$. 
\begin{equation}\label{Mx_rate}
Mx_{rate} = [T_{value}|F_{i}|F_{j}|Sig_{contract}]
\end{equation}
Rating contract evaluates the reputation score $RS_j$ of an \textit{actor j} by updating his previous reputation score $RS'_j$ based on the current feedback $F_j$ and tuning factor $\alpha$ using Eq. \ref{rating}. $RS_i$ is the reputation score of \textit{actor i}. The exponential factor is the aging factor of the impact of contract violation. $\alpha$ is calculated using Eq. \ref{alpha} where $FC$ is the Feedback Credibility (Eq. \ref{FC}), $TV$ is the transaction value (Eq. \ref{TV}) and $CA$ is the collusive activity (Eq. \ref{CA}). $a$ is a positive integer and used to contrast possible collusive actor behaviors.

\begin{equation}\label{rating}
    RS_j = ((1 - \alpha) RS'_j + \alpha F_j) e^{-\frac{c_{fail}}{c_{total}}}
\end{equation}
\begin{equation}\label{alpha}
    \alpha = (FC+TV)CA
\end{equation}
\begin{equation}\label{FC}
   FC = \frac {R_i}{R_i+R_j}.(\frac {F^+}{F^-+F^+})
\end{equation}
\begin{equation}\label{TV}
   TV = \frac {T_{value}} {v_m}
\end{equation}
\begin{equation}\label{CA}
   CA = (\frac {1} {T_{i,j}})^a
\end{equation}

\noindent$\alpha$ tunes the latest received feedback based on the actors' trading history and helps in mitigating malicious activity by a dishonest actor in the marketplace, as explained below.

\noindent \textemdash\ Dishonest feedback: A dishonest actor with a low reputation score may malign other actor's reputation score by giving negative feedback. $FC$ ensures that the feedback credibility of such an actor is low.

\noindent \textemdash\ Ballot stuffing attack: An actor can exploit his high reputation score by giving false negative feedback to other actors. $FC$ mitigates this activity by increasing feedback credibility based on the number of negative feedback ratings given by the actor.

\noindent \textemdash\ Strike and recharge attack: An actor may build up a good reputation by executing several low-value transactions and misbehaving in a very high-value interaction. $TV$ prevents this attack by slightly increasing the reputation score for a low-valued transaction but considerably for a high-valued transaction.

\noindent \textemdash\ Collusion: Two actors may collude with one another to increase each other's reputation score. With an increasing number of transactions between two actors, $CA$ tends to zero. Consequently, feedback loses its relevance.

The above evaluation of the reputation score is suitable for successful transactions only. We also consider the case when the agreement is violated and not fulfilled. Agreement violation should reduce the actor's reputation score drastically, hence discouraging the actor from performing any such violations. To this end, we define a Violation Factor ($VF$) that solely depends on the actor's ability to fulfill the signed contract completely. It holds maximum weight in the calculation of the actor's reputation score as given in Eq. \ref{VF} where ($C_{fail}$) and ($C_{total}$) are the number of failed and total contracts, respectively. The reputation score is calculated using Eq. \ref{RS} as $\alpha$ and $F_j$ are 0 during agreement violation.

\begin{equation}\label{VF}
    VF = 2-2^{(\frac {C_{fail}}{C_{total}})}
\end{equation}

\begin{equation}\label{RS}
    RS_j=RS_j'.VF
\end{equation}

Figure~\ref{smart_contract} depicts how these contracts interact with each other to automate the execution flow of the various stages that are part of data trading. When a seller and a buyer reach an agreement, a new DSC is compiled and deployed on the blockchain. Seller issues $Tx_{add}$ to add subscription details based on their negotiated terms. Seller notifies the buyer and after verifying the terms in DSC, the buyer and seller both use $Tx_{create}$ to register the DSC. Subsequently, $Mx_{price}$ updates the price of the data. At the subscription start time, the seller issues $Tx_{execute}$ to change the status of subscription to active and it returns a session key to the buyer, which is used to encrypt all data in transit to ensure data integrity. Then, data is sent to the buyer and after verifying the data, sends an acknowledgment to the seller. Both parties use a metering system to count the transmission of $N$ data samples to issue $Tx_{settle}$. Based on the count information, DSC verifies if there is a conflict in the count. If there is no dispute, DSC issues an invoice to the buyer and payment to the seller. If not, then the dispute is lodged and payment is not processed. The entities also send their feedback as an input to $Tx_{settle}$. DSC issues $Mx_{rate}$ to the rating contract, which uses respective feedback to update the buyer and seller's reputation scores.

\section{Implementation and Evaluation}

In this section, we first provide a proof-of-concept implementation of the smart contracts, discussed in Section V, using the Solidity version 0.5.12. Then, a single buyer and seller were thoroughly tested on private Ethereum blockchain Ganache. This ensures that the logical flow and execution outcome from the interactions of different smart contracts are as expected. Then we analyze the cost of trade on an Ethereum public testnet. Next, we implement the NR-based algorithm, presented in Section IV, in MATLAB to solve the EDSA problem. The simulation results provide insights into the seller's revenue generation for different scenarios.

\subsection{Proof of concept implementation}

\begin{table}[b!]
  \begin{center}
    \caption{Cost of operations in Ether and USD.}
    \label{table1}
    \begin{tabular}{|c|c|c|c|} 
    \hline
      \textbf{Operations} & \textbf{Gas used} & \textbf{Cost Ethers} & \textbf{Cost (USD)}\\
      \hline
        DSC deployment & 1212806 & 0.0121281 & 1.4675 \\
        $Tx_{create}$ & 86458 & 0.0008646 & 0.10462\\
        $Tx_{add}$ & 265792 & 0.0026579 & 0.32161 \\
        $Tx_{start}$ & 43581 & 0.0004358 & 0.05273\\
        $Tx_{settle}$ & 579676 & 0.0057968 & 0.70141\\
        $Tx_{delete}$ & 22883 & 0.0002288 & 0.02768 \\
      \hline
    \end{tabular}
\end{center}
\end{table}

\begin{table*}[t!]
  \begin{center}
    \caption{Evaluation Parameters.}
    \label{table3}
    \begin{tabular}{|l|c|} 
    \hline
      \textbf{Parameter} & \textbf{Value}\\
      \hline
      Number of IoT devices (i) per seller & $D = 5$\\
      Data type (j) for each device & $S = 5$ \\
      Price of $j^{th}$ type on $i^{th}$ device & $P_{ij}\sim \mathcal{N}(10,\,3)$\\
      Quality of $j^{th}$ type on $i^{th}$ device & $Q_{ij} \sim \mathcal{U}\{10,20,...,100\}$ \\
      Battery of $i^{th}$ device (mAh) & $B_i = 2000$\\
      Number of buyers (k) & $C = 10$\\
      Total number of demands & $dem = 50$\\
      Duration (hr) of $k^{th}$ buyer demand for $j^{th}$ type & $d^k_j \sim \mathcal{N}(5,\,2)$ \\
      Sampling interval (min) of $k^{th}$ buyer demand for $j^{th}$ type & $s^k_j\sim \mathcal{N}(10,\,60)$\\
      Quality demanded by $k^{th}$ buyer & $q^k \sim \mathcal{U}\{10,20,...,100\}$\\
      \hline
    \end{tabular}
  \end{center}
\end{table*}

The implementation was done on the Ropsten testnet network, the public Ethereum network, which behaves similarly to a production blockchain. It runs a proof-of-work consensus and used for testing purposes. Our smart contracts\footnote{Smart contract codes available at https://github.com/pooja239/DataMart} were implemented and deployed using Remix IDE\footnote{http://remix.ethereum.org/}.

It is worth noting that any operation or transaction that modifies the state incurs fees, which need to be paid by the involved parties. These costs are estimated using the amount of gas consumed and the unit gas price. The gas consumed during any operation reflects the computational complexity or size of the smart contracts, while the miners in the system determine the gas prices. We used 10 Gwei gas price to evaluate the cost of different operations during data trading (the recommended gas prices can be found at \cite{b29}). Table~\ref{table1} shows the costs of DSC contract deployment and the various transactions to achieve end-to-end data trading, where the total cost of all the operations sums up be USD 2.67555.

It can be observed that $Tx_{add}$ and $Tx_{settle}$ consumed more gas than other transactions. This is because these transactions involve write operations on the blockchain. $Tx_{add}$ requires interaction with the pricing contract to fetch the price of the subscription and also update the pricing information in the ledger. At the same time, $Tx_{settle}$ interacts with the rating contract to calculate and update the reputation score based on the latest feedback. It can be observed that deploying a new contract is expensive. Thus, a deployed trade contract could be reused if it provides a satisfactory trading experience for the users.

\subsection{Evaluation}

Table~\ref{table3} gives the values of the parameters, defined in section IV, used in the simulations unless stated otherwise. The performance results are generated by averaging results over 1000 iterations. 

\begin{figure}[b!]
\centerline{\includegraphics[width=\linewidth]{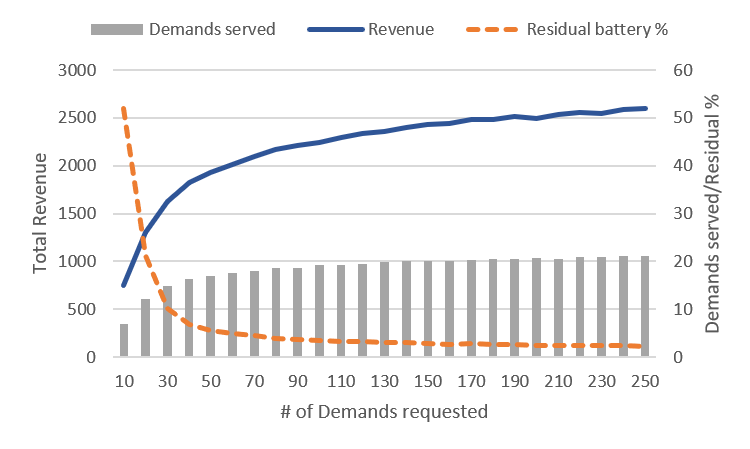}}
\caption{Revenue generation and demand selection with NR-based algorithm}
\label{fig5}
\end{figure}

Figure~\ref{fig5} shows the effect of increasing the number of demands requested for 10 data types by 50 buyers. We can observe that increasing the number of demands requested from 10 to 250 causes a logarithmic growth for the revenue, which reaches saturation at a certain demand level (210-230) due to the limited batteries of the devices. This can be observed from the residual battery levels. Note that buyers' quality requirements may also limit the number of demands served by the seller, as some demands may have higher quality requirements than the quality of data provided by the seller.

\begin{figure}[b!]
\centerline{\includegraphics[width=\linewidth]{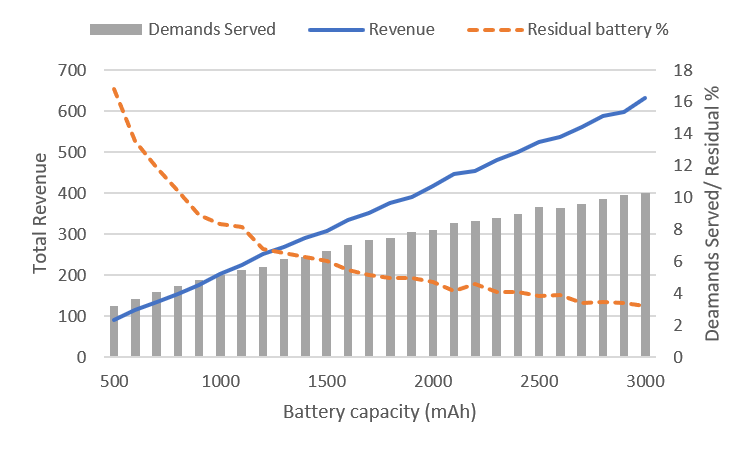}}
\caption{Effect of increasing battery capacity}
\label{fig6}
\end{figure}

Next, Figure~\ref{fig6} demonstrates the effect of increasing the battery capacity on the total revenue of a seller with a single device. We observe that the revenue increases linearly with the increasing battery capacity from 500mAh to 3000mAh. As the battery capacity increases, the seller can select and serve more demands and generates higher revenue. The residual battery does not change much; however, the percentage of the residual battery decreases, resulting in better battery capacity utility. It is observed that with the linear increment of battery capacity, the total revenue generated is increasing linearly.

\begin{figure}[t!]
\centerline{\includegraphics[width=\linewidth]{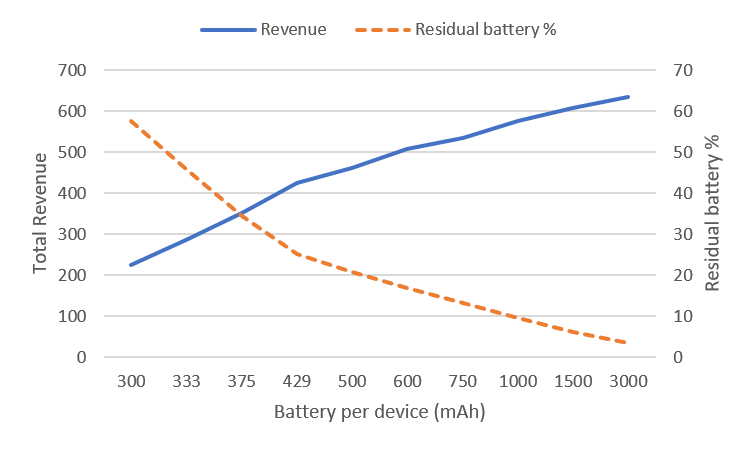}}
\caption{Effect of increasing number of devices}
\label{fig7}
\end{figure}

Finally, we analyze the impact of increasing the number of seller's devices on the generated revenue in Figure~\ref{fig7}. The overall battery capacity with a seller is kept fixed to capture the original trend of increasing the number of devices rather than increasing battery which will linearly increase the generated revenue. We observe that increasing battery capacity for IoT devices is more strongly related to revenue generation than increasing the number of devices. As we move from 10 devices with 300mAh batteries to 1 device with 3000mAh battery, the total revenue increases. This could be explained by the smaller battery capacities and residual batteries remaining on the devices. Devices with small batteries cannot serve demands that require high energy consumption. Furthermore, the residual batteries remaining on the devices cannot be combined and utilized to serve more demands.

\section{Conclusion}
We have presented the design of a decentralized marketplace equipped with essential components such as data agreement management, pricing model and rating mechanism. We presented preliminary work showing how to implement these components using smart contracts. Also, we evaluate the gas consumption and cost incurred for interacting with the proposed marketplace framework. Furthermore, we formulated the EDSA problem that maximizes a user's utility to improve the sustainability and usability of the marketplace. We simulated various scenarios to analyze the impact of the requested demands on the battery drainage of the seller's devices. In our future work, we will consider maximizing the number of demands selected in a multi-objective optimization for demand selection and allocation problem. We will also analyze the performance of the proposed real-time IoT data marketplace and the NR-based algorithm experimentally using IoT hardware.


\begin{thebibliography}{00}
\bibitem{b1} Mišura, Krešimir, and Mario Žagar. "Data marketplace for Internet of Things." 2016 International Conference on Smart Systems and Technologies (SST). IEEE, 2016.
\bibitem{b2} Cao, Tien-Dung, et al. "MARSA: A marketplace for realtime human sensing data." ACM TOIT 16.3 (2016): 16.
\bibitem{b3} Zheng, Zibin, et al. "Blockchain challenges and opportunities: A survey." International Journal of Web and Grid Services 14.4, 2018, pp. 352-375.
\bibitem{b4} Xiao Y, Zhang N, Lou W, Hou YT. "Enforcing Private Data Usage Control with Blockchain and Attested Off-chain Contract Execution." arXiv preprint arXiv:1904.07275, 2019.
\bibitem{b5} S. Bajoudah, C. Dong and P. Missier, "Toward a Decentralized, Trust-Less Marketplace for Brokered IoT Data Trading Using Blockchain," 2019 IEEE International Conference on Blockchain (Blockchain), Atlanta, GA, USA, 2019, pp. 339-346.
\bibitem{b6} Missier, Paolo, et al. "Mind my value: a decentralized infrastructure for fair and trusted iot data trading." Proceedings of the Seventh International Conference on the Internet of Things. 2017.
\bibitem{b7} Mao, Weichao, Zhenzhe Zheng, and Fan Wu. "Pricing for revenue maximization in iot data markets: An information design perspective." IEEE INFOCOM 2019-IEEE Conference on Computer Communications. IEEE, 2019.
\bibitem{b8} Graf, Thomas. "Power-efficient positioning technologies for mobile devices." Berlin University of Technology, SNET2 Seminar. 2011.
\bibitem{b9} X. Wang, A. K. Wong and Y. Kong, "Mobility tracking using GPS, Wi-Fi and Cell ID," The International Conference on Information Network 2012, Bali, 2012, pp. 171-176.
\bibitem{b10} P. Gupta, S. S. Kanhere and R. Jurdak. "A Decentralized IoT Data Marketplace." in Proceedings of 3rd Symposium on Distributed Ledger Technology (SDLT), Gold Coast, November 2018.
\bibitem{b11} Angelakis, Vangelis, Ioannis Avgouleas, Nikolaos Pappas, Emma Fitzgerald, and Di Yuan. "Allocation of heterogeneous resources of an IoT device to flexible services." IEEE Internet of Things Journal 3, no. 5 (2016): 691-700.
\bibitem{b12} Choi, Yeongho, and Yujin Lim. "Optimization approach for resource allocation on cloud computing for iot." International Journal of Distributed Sensor Networks 12.3 (2016): 3479247.
\bibitem{b13} Ghanbari, Zahra, Nima Jafari Navimipour, Mehdi Hosseinzadeh, and Aso Darwesh. "Resource allocation mechanisms and approaches on the Internet of Things." Cluster Computing 22, no. 4 (2019): 1253-1282.
\bibitem{b14} Shafagh, Hossein, Lukas Burkhalter, Anwar Hithnawi, and Simon Duquennoy. "Towards blockchain-based auditable storage and sharing of iot data." In Proceedings of the 2017 on Cloud Computing Security Workshop, pp. 45-50. 2017.
\bibitem{b15} Özyilmaz, Kazim Rifat, Mehmet Doğan, and Arda Yurdakul. "IDMoB: IoT data marketplace on blockchain." 2018 Crypto Valley Conference on Blockchain Technology (CVCBT). IEEE, 2018.
\bibitem{b16} Ramachandran, Gowri Sankar, Rahul Radhakrishnan, and Bhaskar Krishnamachari. "Towards a decentralized data marketplace for smart cities." IEEE International Smart Cities Conference (ISC2). IEEE, 2018.
\bibitem{b17} A. Suliman, Z. Husain, M. Abououf, M. Alblooshi and K. Salah, "Monetization of IoT data using smart contracts," in IET Networks, vol. 8, no. 1, pp. 32-37, 1 2019.
\bibitem{b18} Xu, R., Ramachandran, G.S., Chen, Y., Krishnamachari, B.: Blendsm-ddm: Blockchain enabled secure microservices for decentralized data marketplaces. In: 2019 IEEE International Smart Cities Conference (ISC2). IEEE (2019)
\bibitem{b19} R. Radhakrishnan, G. S. Ramachandran and B. Krishnamachari, "SDPP: Streaming Data Payment Protocol for Data Economy," 2019 IEEE International Conference on Blockchain and Cryptocurrency (ICBC), Seoul, Korea (South), 2019, pp. 17-18.
\bibitem{b20} Mišura, Krešimir, and Mario Žagar. "Negotiation in internet of things." Automatika: časopis za automatiku, mjerenje, elektroniku, računarstvo i komunikacije 57.2 (2016): 304-318.
\bibitem{b21} C. I. Podilchuk and E. J. Delp, "Digital watermarking: algorithms and applications," in IEEE Signal Processing Magazine, vol. 18, no. 4, pp. 33-46, July 2001.
\bibitem{b22} Charpenay, Victor, et al. "Matching offerings and queries on an Internet of Things marketplace." European Semantic Web Conference. Springer, Cham, 2018.
\bibitem{b23} Carroll, Aaron, and Gernot Heiser. "An Analysis of Power Consumption in a Smartphone." USENIX annual technical conference. Vol. 14. 2010.
\bibitem{b24} Dawande, Milind, et al. "Approximation algorithms for the multiple knapsack problem with assignment restrictions." Journal of combinatorial optimization 4.2 (2000): pp.171-186.
\bibitem{b25} Dahl, Geir, and Njål Foldnes. "LP based heuristics for the multiple knapsack problem with assignment restrictions." Annals of Operations Research 146.1 (2006): 91-104.
\bibitem{b26} Murthy, Ananth, Chandan Yeshwanth, and Shrisha Rao. "Distributed Approximation Algorithms for the Multiple Knapsack Problem." arXiv preprint arXiv:1702.00787 (2017).
\bibitem{b27} Molina, Victor, Marta Kersten-Oertel, and Tristan Glatard. "A conceptual marketplace model for iot generated personal data." arXiv preprint arXiv:1907.03047 (2019).
\bibitem{b28} Lax, G. and Sarné, G.M., 2008. CellTrust: a reputation model for C2C commerce. Electronic Commerce Research, 8(4), pp.193-216.
\bibitem{b29} "ETH Gas Station", Ethgasstation.info, 2020. [Online]. Available: https://ethgasstation.info/. [Accessed: 23- Jan- 2020].
\bibitem{b30} "Introduction to Smart Contracts — Solidity 0.6.2 documentation", Solidity.readthedocs.io, 2020. [Online]. Available: https://solidity.readthedocs.io/en/develop/introduction-to-smart-contracts.html\#self-destruct. [Accessed: 08- Jan- 2020].
\end{thebibliography}
\end{document}